\documentclass[12pt,thmsa]{article}
\usepackage{sw20lart}

\input{tcilatex}

\begin{document}

\author{{\large F. Iddir}$\thanks{
E-mail: \emph{iddir@univ-oran.dz}}${\large \ \ }and{\large \ L. Semlala}$%
\thanks{
E-mail: \emph{l\_semlala@yahoo.fr}}$ \\
%EndAName
\textit{Laboratoire\ de Physique\ Th\'{e}orique, Universit\'{e} d}'\textit{%
Oran Es-S\'{e}nia, }\\
\textit{31100, } \textit{ALGERIA}}
\date{03 November 2004}
\title{{\LARGE Quarks-excited states or gluon-excited states, are the J}$%
^{PC}=${\LARGE 1}$^{-+}${\LARGE \ hybrid mesons confirmed?}}
\maketitle

\begin{abstract}
The existence of hadrons containing gluons as constituent particles (like
hybrid mesons) will involve the necessity to admit an hadronic spectroscopy
beyond the na\"{\i}ve quark model. In the QGC model, the lowest J$^{PC}$= 1$%
^{-+}\ $hybrid mesons may be built in two different modes: the GE-mode
corresponding to an angular momentum between the gluon and $q\overline{q}\ $%
system, while the QE-mode corresponding to an angular momentum between\ q
and\ $\overline{q}$. In this paper, we give an analysis of the possible $%
1^{-+}$ hybrid mesons in the 1.32-2.04 GeV mass range, and predictions on
their decay. We discuss on the inconsistency between some theoretical
results and experimental\ data.We find that the 1$^{-+}$(1400\ MeV),
dominated by the QE-mode will decay in the favourite $\rho \pi $ channel,
while the GE-1$^{-+}$(1600\ MeV), and the 1$^{-+}$(2.0\ GeV), pure
GE-hybrids, will decay preferably into $b_{1}\pi $ channel (never observed);
we indicate problems concerning the observation of decay channels which are
forbidden by theoretical present models.The total decay widths of the 1$%
^{-+} $(2.0\ GeV) and the GE-1$^{-+}$(1600\ MeV) being very large, this
leads to a problem of their observability as resonances.

We may conclude that at this time, there is no absolute confirmation of
their existence, but some asked questions remain not solved.
\end{abstract}

\section{Introduction}

The na\"{\i}ve quark model interprets the meson as a bound state of a quark
and an antiquark, and predicts that the charge conjugation $(C)$ and the
parity $(P)$ of a meson with spin $S$ and angular momentum $L$ are
respectively: $C=(-)^{L+S}$ and $P=(-)^{L+1}$.

Besides these conventional mesons, Quantum Chromodynamics, actually
considered as the theory of strong interactions, predicts that states
containing gluons (namely glueballs and hybrids) may exist. Several models
(Lattice-QCD, QCD Sum-rules, Flux Tube Model, Constituent Glue Model, ...
suggest that lightest glueballs occur in the 1-2\ GeV region and $q\overline{%
q}g$ hybrids occur around 2\ GeV in mass, but without any clear evidence.

The collective behaviour of gluons in the strong interaction regime of QCD
is not yet established, and the confirmation of presence of gluons in
hadrons as constituent particles will change our understanding about the r%
\^{o}le of the gluons in QCD. From the theoretical side such state may be in
colorsinglet; however this does not imply the existence of resonance: one
must demonstrate a state with consider quantum numbers and which has the
total decay width smaller compared to the level spacing (if the total decay
width is larger than the level spacing, the considered is simply a part of
the non-resonant background) and a long life time before decaying.

The most clear-out signal for hybrid meson is to reach for $J^{PC}$ quantum
numbers not allowed in the na\"{\i}ve quark model, such as $%
J^{PC}=0^{+-},0^{--},1^{-+},\ 2^{+-},$ ... There has been considerable
recent interest in such hybrid mesons, both from the perspective of models
and from the experimental data.

From experimental efforts at IHEP, KEK, CERN, and BNL, hybrid $1^{-+}$
candidates have been observed by AGS at BNL\cite{Thomp, Adams, Chung, Lee,
Ivanov}, by Cristall Barrel at LEAR\cite{Abele}, by GAMS at CERN\cite{Alde},
by VES at IHEP\cite{Bela, Proko}, by FNAL\cite{Black}, ... Furthermore,
several experimental programs are proposed for further investigation of
hybrids. COMPASS Program at CERN\cite{Moinester} intends to study the
possibility to Primakov production of hybrids near 1.4 GeV; hybrids with $%
J^{PC}=0^{+-},2^{+-}$ and $1^{-+}$ would be produced by photoproduction at
CEBAF\cite{Barnes} and $2^{++}$ at JLAB\cite{Achasov}. It is proposed to
investigate mainly the $J^{PC}=$ $1^{-+}$ : programs are proposed to search
for resonances decaying in b$_{1}\pi $ (at SLAC\cite{Grb}) and $\rho \pi $
(at BEPC/VES \cite{Lie}, and by photoproduction \cite{Afana}). Then, this
intense experimental activity should contribute significantly to the
understanding of the hybrids, and would give answer to the question of their
existence.

In this paper, we give an analysis of the different hybrid modes possible to
construct in the 1.32-2.04 GeV mass range and we give predictions on their
decay. We discuss on the inconsistency between some theoretical results and
experimental data, concerning the favourite channels (non observed) and
observed channels (non allowed); we suggest to attribute a particular
importance to the problem of the existence of hybrid mesons, since earlier
theoretical results lead to this interrogation. \ In a preceding work\cite%
{Safir1}, we have studied the states of the $J^{PC}=$ $1^{--}$ $c\overline{c}%
g$ charmonium hybrid with a mass around 4 GeV; due to the very large total
decay width, we concluded that the $c\overline{c}g$ at 4 GeV cannot be a
resonance and then not a physical observable object.

Furthermore, the study of the $J^{PC}=$ $1^{-+}$ hybrid candidates with
masses 1.4 and 1.6 GeV\cite{Safir2} give b$_{1}\pi $ as preferred decay
channel (which is not observed) and $\eta \pi $ forbidden (which is
observed). We give also in this work our interpretation of the exotic 2.0
GeV resonance claimed to be observed by the AGS Collaboration at BNL\cite%
{Lee}, decaying in f$_{1}\pi $.

\section{The two-modes hybrid mesons and their masses}

In the quark-gluon constituent model\cite{LeYaouanc, Iddir}, the gluon is
considered as a massive constituent particle ($m_{g}\simeq 800MeV$) moving
in the framework of $q\overline{q}$ pair center of mass. Restricting
ourselves to the lightest states, the hybrid $1^{-+}$ states may be built in
two ways: with $l_{q\overline{q}}=1$, $S_{q\overline{q}}=0$, $l_{g}=0$
(which is the quarks-excited mode: QE) and with $l_{q\overline{q}}=0$, $S_{q%
\overline{q}}=1$, $l_{g}=1$ (which is the gluon-excited mode: GE); we
examine also the possibility to have a mixed state (QE+GE). The calculations
of the hybrid masses, using two different confining potentiel models to
describe the hybrid states, show a difference between the masses of the two
modes, the GE-hybrid being much heavier:

- a non-relativistic quark model with a confining chromo-harmonic potential 
\cite{These Sem}

- a more realistic potential in good agreement with QCD characteristics
(with coulombien and linear terms), and taking in account some relativistic
corrections\cite{Semlala}.

N.B: We note the quantum numbers of the hybrid meson:

$\emph{l}_{g}$: is the relative orbital momentum of the gluon in the \emph{q%
\={q}} center of mass;

$\emph{l}_{\text{\textit{q\={q}}}}\emph{\ \ }$: is the relative orbital
momentum between \emph{q} and \emph{\={q}};

\emph{S}$_{\text{\textit{q\={q}}}}$\emph{\ }: is the total quark spin;

\emph{j}$_{\text{\textit{g }}}$\emph{\ \ }: is the total gluon angular
momentum;

\emph{L \ \ }: \textit{l}$_{\text{\textit{q\={q}}}}$ + \textit{j}$_{\text{%
\textit{g}}}.$

The parity and charge conjugation of the hybrid are given by: 
\begin{equation}
\begin{array}{l}
P=\left( -\right) ^{l_{q\bar{q}}+l_{g}}; \\ 
C=\left( -\right) ^{l_{q\bar{q}}+S_{q\bar{q}}+1}.%
\end{array}
\tag{1}
\end{equation}

Wave functions and masses have been calculated through the equation:

\begin{equation}
\left\{ \sum\limits_{i=1}^{N}\left( \frac{\vec{p}_{i}^{\text{ }2}}{2M_{i}}+%
\frac{M_{i}}{2}+\frac{m_{i}^{2}}{2M_{i}}\right) +V_{eff}\right\} \Psi (\vec{r%
}_{i})=E\Psi (\vec{r}_{i});  \tag{2}
\end{equation}

which take in account relativistic corrections in the Hamiltonian; $V_{eff}$
is the average over the color space of chromo-spatial potential\cite{Semlala}%
:

\begin{eqnarray}
V_{eff} &=&\left\langle V\right\rangle _{color}=\left\langle
-\sum\limits_{i<j=1}^{N}\mathbf{F}_{i}\cdot \mathbf{F}_{j}\text{ }%
v(r_{ij})\right\rangle _{color}  \TCItag{3} \\
&=&\sum\limits_{i<j=1}^{N}\alpha _{ij}v(r_{ij})\text{ };  \nonumber
\end{eqnarray}%
where $v(r_{ij})$ is the phenomenological potential term, having a
(Coulomb+linear) or harmonic oscillator form.

Defining the Jacobi coordinates:

\[
\begin{array}{l}
\vec{\rho}=\vec{r}_{\bar{q}}-\vec{r}_{q}; \\ 
\vec{\lambda}=\vec{r}_{g}-\frac{M_{q}\vec{r}_{q}+M_{\bar{q}}\vec{r}_{\bar{q}}%
}{M_{q}+M_{\bar{q}}}.%
\end{array}%
\]

We develop the spatial wave functions:%
\begin{equation}
\psi ^{l_{q\bar{q}}l_{g}}(\vec{\rho},\vec{\lambda})=\sum%
\limits_{n=1}^{N}a_{n}\varphi _{n}^{l_{q\bar{q}}l_{g}}(\vec{\rho},\vec{%
\lambda})\text{ };  \tag{4}
\end{equation}%
where $\varphi _{n}^{l_{q\bar{q}}l_{g}}(\vec{\rho},\vec{\lambda})$ are the
Gaussian-type functions.

\subsection{\protect\bigskip Masses of the GE- and QE-hybrids}

We present in Table 1 estimates of hybrid mesons masses, using the harmonic
oscillator confining potential and the (Coulomb + Linear) potential, for
both light and heavy flavors; we take $800MeV$ for the mass of the gluon.

\begin{center}
\begin{tabular}{|l|l|l|l|l|}
\hline
potentiel model \ \ \ \ \ \ \ \ \ \ \ \ \  & $\ u,d$ & $\ \ \ s$ & $\ \ \ c$
& \ \ \ $b$ \\ \hline
Coul.+Lin.\ \ \ \ \ \ \ \ \ \ .$%
\begin{array}{c}
\text{QE} \\ 
\text{GE}%
\end{array}
$ & $%
\begin{array}{c}
1.31 \\ 
1.70%
\end{array}
$ & $%
\begin{array}{c}
1.57 \\ 
2.00%
\end{array}
$ & $%
\begin{array}{c}
4.09 \\ 
4.45%
\end{array}
$ & $%
\begin{array}{c}
10.34 \\ 
10.81%
\end{array}
$ \\ \hline
Harm.Oscill. $\ \ \ \ \ \ \ 
\begin{array}{c}
\text{QE} \\ 
\text{GE}%
\end{array}
$ & $%
\begin{array}{c}
1.47 \\ 
1.61%
\end{array}
$ & $%
\begin{array}{c}
1.68 \\ 
1.84%
\end{array}
$ & $%
\begin{array}{c}
4.83 \\ 
4.98%
\end{array}
$ & $%
\begin{array}{c}
11.40 \\ 
11.58%
\end{array}
$ \\ \hline
\end{tabular}

\bigskip \textit{Table 1: estimates of hybrid mesons masses (in GeV) for
different flavors (without spin effects)}
\end{center}

In Table 2, we give results of light hybrid mesons masses, calculated using
the (Coulomb + Linear) potential, and taking in account spin-spin effects.

\[
\begin{tabular}{|c|c|c|c|c|}
\hline
& $u\bar{u}g$ & $u\bar{s}g$ & $s\bar{s}g$ &  \\ \hline
\begin{tabular}{l}
$S=1$%
\end{tabular}
& 
\begin{tabular}{l}
$1.32$%
\end{tabular}
& 
\begin{tabular}{l}
$1.45$%
\end{tabular}
& 
\begin{tabular}{l}
$1.58$%
\end{tabular}
& 
\begin{tabular}{l}
QE Mode ($l_{q\bar{q}}=1;$ $l_{g}=0$ and $S_{q\bar{q}}=0$)%
\end{tabular}
\\ \hline
\begin{tabular}{l}
$S=0$%
\end{tabular}
& 
\begin{tabular}{l}
$1.56$%
\end{tabular}
& 
\begin{tabular}{l}
$1.72$%
\end{tabular}
& 
\begin{tabular}{l}
$1.87$%
\end{tabular}
&  \\ \hline
$%
\begin{array}{c}
S=1%
\end{array}
$ & $%
\begin{array}{c}
1.69%
\end{array}
$ & $%
\begin{array}{c}
1.84%
\end{array}
$ & $%
\begin{array}{c}
1.99%
\end{array}
$ & 
\begin{tabular}{l}
GE Mode ($l_{q\bar{q}}=0;$ $l_{g}=1$ and $S_{q\bar{q}}=1$)%
\end{tabular}
\\ \hline
$%
\begin{array}{c}
S=2%
\end{array}
$ & $%
\begin{array}{c}
1.75%
\end{array}
$ & $%
\begin{array}{c}
1.89%
\end{array}
$ & $%
\begin{array}{c}
2.04%
\end{array}
$ &  \\ \hline
\end{tabular}
\]

\begin{center}
\textit{Table 2: masses of }$1^{-+}$\textit{\ light hybrid mesons (in GeV),
calculated (using Cb+Lin. pot.) within spin-spin corrections}\cite{Isgur}
\end{center}

\subsection{\protect\bigskip Masses of the mixed (GE+QE) states}

Representing the hybrid meson wave function in the cluster approximation\cite%
{Semlala}:

\begin{eqnarray}
\Psi _{JM}(\vec{\rho},\vec{\lambda}) &=&\sum\limits_{n,\text{ }l_{q\bar{q}},%
\text{ }l_{g}}a_{n}^{l_{q\bar{q}}l_{g}}\sum\limits_{j_{g},\text{ }L,\text{ (}%
m),\text{ (}\mu )}\varphi _{n}^{l_{q\bar{q}}l_{g}}(\vec{\rho},\vec{\lambda})%
\text{\textbf{e}}^{\mu _{g}}\chi _{_{S_{_{q\bar{q}}}}}^{^{\mu _{q\bar{q}%
}}}\left\langle l_{g}m_{g}1\mu _{g}\mid J_{g}M_{g}\right\rangle  \TCItag{5}
\\
&&\times \left\langle l_{q\bar{q}}m_{q\bar{q}}J_{g}M_{g}\mid Lm\right\rangle
\left\langle LmS_{q\bar{q}}\mu _{q\bar{q}}\mid JM\right\rangle \text{ .} 
\nonumber
\end{eqnarray}

the expansion of the wave function for a mixed (GE+QE) state may be written:

\begin{equation}
\Psi _{1^{-\text{ }+}}(\vec{\rho},\vec{\lambda})\simeq
\sum\limits_{n=1}^{N}a_{n}^{QE}\varphi _{n}^{QE}(\vec{\rho},\vec{\lambda}%
)+\sum\limits_{n=1}^{N}a_{n}^{GE}\varphi _{n}^{GE}(\vec{\rho},\vec{\lambda}).
\tag{6}
\end{equation}

For the spin states we choused $\left\{ \left| S_{q\bar{q}},\text{ }s_{g};%
\text{ }S\right\rangle \right\} $ ( $s_{g}=1$ and $\mathbf{S}=\mathbf{S}_{q%
\bar{q}}+\mathbf{s}_{g}$ ).\linebreak

Using (Cb+Lin.) confining potential, we give the results in Table 3:

\begin{center}
\begin{tabular}{|c|c|c|}
\hline
hybrid state & 2-modes mixing & mass (in GeV) \\ \hline
$\left\vert 1^{-\text{ }+}(u\bar{u}g)\right\rangle $ & $-.999\left\vert
QE\right\rangle +.040\left\vert GE\right\rangle $ & $1.34\text{ }$ \\ \hline
$\left\vert 1^{-\text{ }+}(u\bar{u}g)\right\rangle $ & $-\left\vert
GE\right\rangle $ & $1.72\text{ }$ \\ \hline
$\left\vert 1^{-\text{ }+}(s\bar{s}g)\right\rangle $ & $-.999\left\vert
QE\right\rangle +.050\left\vert GE\right\rangle $ & $1.60\text{ }$ \\ \hline
$\left\vert 1^{-\text{ }+}(s\bar{s}g)\right\rangle $ & $-\left\vert
GE\right\rangle $ & $2.02\ $ \\ \hline
$\left\vert 1^{-\text{ }+}(c\bar{c}g)\right\rangle $ & $-.999\left\vert
QE\right\rangle -.040\left\vert GE\right\rangle $ & $4.10\text{ }$ \\ \hline
$\left\vert 1^{-\text{ }+}(c\bar{c}g)\right\rangle $ & $-.031\left\vert
QE\right\rangle -.999\left\vert GE\right\rangle $ & $4.45\text{ }$ \\ \hline
\end{tabular}

\textit{Table 3: predictions of mixed (QE+GE) states}
\end{center}

The results show that the QE-hybrid and the GE-hybrid mix very weakly.

\section{The decay of the two-modes hybrid mesons}

We use the quark-gluon constituent model, in which the gluon moves in the
framework of the $q\overline{q}$ pair center of mass. It is important to
note that the decay occurs through two diagrams, the gluon annihilating into
a $q\overline{q}$ pair (namely $q=u,d$ and $s$) and the decay amplitude will
then be the sum of two corresponding amplitudes to each contribution\cite%
{Safir1, Safir2}.

Then the decay of an hybrid state A into two mesons B and C is represented
by the matrix element of the Hamiltonian annihilating a gluon and creating a
quark pair:

\begin{equation}
\langle BC\left\vert H\right\vert A\rangle =gf(A,B,C)\left( 2\pi \right)
^{3}\delta _{3}\left( p_{A}-p_{B}-p_{C}\right) ;  \tag{7}
\end{equation}

where $f(A,B,C)$ representing the decay amplitude by:

\begin{eqnarray}
f(A,B,C) &=&\sum_{\left( m\right) ,\left( \mu \right) }\Phi \Omega X\left(
\mu _{q\overline{q}},\mu _{g};\mu _{B},\mu _{C}\right) I\left( m_{q\overline{%
q}},m_{g};m_{B},m_{C},m\right)  \TCItag{8} \\
&&\times \left\langle l_{g}m_{g}1\mu _{g}\right\vert J_{g}M_{g}\rangle
\langle l_{q\overline{q}}m_{q\overline{q}}J_{g}M_{g}\left\vert Lm^{\prime
}\right\rangle \langle Lm^{\prime }S_{q\overline{q}}\mu _{q\overline{q}%
}\left\vert JM\right\rangle  \nonumber \\
&&\times \langle l_{B}m_{B}S_{B}\mu _{B}\left\vert J_{B}M_{B}\right\rangle
\langle l_{C}m_{C}S_{C}\mu _{C}\left\vert J_{C}M_{C}\right\rangle . 
\nonumber
\end{eqnarray}

where $\Phi ,\Omega ,X$ and $I$ are the flavor, color, spin and spatial
overlaps. $\Omega $ is given by:

\begin{equation}
\Omega =\frac{1}{24}\sum_{a}tr\left( \lambda ^{a}\right) ^{2}=\frac{2}{3}. 
\tag{9}
\end{equation}

From:

\begin{equation}
\chi _{\mu _{1}}^{+}\sigma ^{\lambda }\chi _{\mu _{2}}=\sqrt{3}\langle \frac{%
1}{2}\mu _{2}1\lambda \left\vert \frac{1}{2}\mu _{1}\right\rangle ,  \tag{10}
\end{equation}

we obtain the spin overlap:

\begin{eqnarray}
X\left( \mu _{q\overline{q}},\mu _{g};\mu _{B},\mu _{C}\right) &=&\sum_{S}%
\sqrt{2}\left[ 
\begin{array}{ccc}
1/2 & 1/2 & S_{B} \\ 
1/2 & 1/2 & S_{C} \\ 
S_{q\overline{q}} & 1 & S%
\end{array}%
\right]  \TCItag{11} \\
&&\times \langle S_{q\overline{q}}\mu _{q\overline{q}}1\mu _{g}\left\vert
S\left( \mu _{q\overline{q}}+\mu _{g}\right) \right\rangle \langle S_{B}\mu
_{B}S_{C}\mu _{C}\left\vert S\left( \mu _{B}+\mu _{C}\right) \right\rangle ;
\nonumber
\end{eqnarray}

where

\[
\left[ 
\begin{array}{ccc}
1/2 & 1/2 & S_{B} \\ 
1/2 & 1/2 & S_{C} \\ 
S_{q\overline{q}} & 1 & S%
\end{array}
\right] =\sqrt{3\left( 2S_{B}+1\right) \left( 2S_{C}+1\right) \left( 2S_{q%
\overline{q}}+1\right) }\left\{ 
\begin{array}{ccc}
1/2 & 1/2 & S_{B} \\ 
1/2 & 1/2 & S_{C} \\ 
S_{q\overline{q}} & 1 & S%
\end{array}
\right\} . 
\]

The spatial overlap is given by:

\begin{eqnarray}
I\left( m_{q\overline{q}},m_{g};m_{B},m_{C},m\right) &=&\iint \frac{d%
\overrightarrow{p}d\overrightarrow{k}}{\left( 2\pi \right) ^{6}\sqrt{2\omega 
}}\Psi _{q\overline{q}g}^{l_{q\overline{q}}m_{q\overline{q}%
}l_{g}m_{g}}\left( \overrightarrow{P}_{B}-\overrightarrow{p},\overrightarrow{%
k}\right)  \TCItag{12} \\
&&\times \Psi _{q_{i}\overline{q}}^{l_{B}m_{B}\ \ast }\left( \overrightarrow{%
p}_{1}\right) \Psi _{q\overline{q}_{i}}^{l_{C}m_{C}\ \ast }\left( 
\overrightarrow{p}_{2}\right) Y_{l}^{m\ \ast }\left( \Omega _{B}\right)
d\Omega _{B},  \nonumber
\end{eqnarray}

where:

\begin{equation}
\overrightarrow{p}_{1}=\frac{m_{\overline{q}_{i}}}{m_{q}+m_{\overline{q}_{i}}%
}\overrightarrow{P}_{B}-\overrightarrow{p}-\frac{\overrightarrow{k}}{2} 
\tag{13}
\end{equation}

\begin{equation}
\overrightarrow{p}_{2}=-\frac{m_{q_{i}}}{m_{\overline{q}}+m_{q_{i}}}%
\overrightarrow{P}_{B}+\overrightarrow{p}-\frac{\overrightarrow{k}}{2}. 
\tag{14}
\end{equation}

$l,m\ \ $label the orbital momentum between the two final mesons.

Finally,

\begin{equation}
\Phi =\left[ 
\begin{array}{ccc}
i_{1} & i_{3} & I_{B} \\ 
i_{2} & i_{4} & I_{C} \\ 
S_{q\overline{q}} & 1 & I_{A}%
\end{array}%
\right] \eta \epsilon ;  \tag{15}
\end{equation}

where I's (i's) label the hadron (quark) isospins, $\eta =1$ if the gluon
goes into strange quarks and $\eta =\sqrt{2}$ if it goes into non strange
ones. $\epsilon $ is the number of diagrams contributing to the decay.
Indeed one can check that since P and C are conserved, two diagrams
contribute with the same sign and magnitude for allowed decays while they
cancel for forbidden ones. In the case of two identical final particles, $%
\epsilon =\sqrt{2}$.

The partial width is then given by:

\begin{equation}
\Gamma \left( A\rightarrow BC\right) =4\alpha _{s}\left\vert f\left(
A,B,C\right) \right\vert ^{2}\frac{P_{B}E_{B}E_{C}}{M_{A}};  \tag{16}
\end{equation}

with

\begin{equation}
P_{B}^{2}=\frac{\left[ M_{A}^{2}-\left( m_{B}+m_{C}\right) ^{2}\right] \left[
M_{A}^{2}-\left( m_{B}-m_{C}\right) ^{2}\right] }{4M_{A}^{2}};  \tag{17}
\end{equation}

\begin{eqnarray}
E_{B} &=&\sqrt{P_{B}^{2}+m_{B}^{2}};  \TCItag{18} \\
E_{B} &=&\sqrt{P_{B}^{2}+m_{C}^{2}}.  \nonumber
\end{eqnarray}

\subsection{The Quarks-Excited mode, decaying in $\protect\rho \protect\pi $}

Using the conservation of angular momentum, parity, isospin and G-parity for
the state with :

\[
l_{q\overline{q}}=1,\ l_{g}=0,\ S_{q\overline{q}}=0,\ L=1\ \text{and }J=1 
\]

the $J^{PC}=1^{-+}$-QE hybrid is allowed to decay into (the decay channels,
with the widths for the $J^{PC}=1^{-+}$ hybrid meson at $1.4$ and $1.6GeV$
are given in \cite{Safir2}) :

\[
\rho \pi ,\ \rho \omega ,\ \rho (1450)\pi ,\ K^{\ast }(1410)K\text{ and }%
K^{\ast }K. 
\]

All this channels verify the selection rule for the $1^{-+}$ QE-hybrid mode
which is allowed to decay only into two fundamental mesons $\left(
L=0+L=0\right) $\cite{Safir1, Safir2}$.$

The preferred decay channel of the $1^{-+}$ QE-hybrid meson is $\rho \left(
770\right) \pi $, and the calculations give a width $\Gamma =296MeV$ (for $%
M=2.0GeV$). The decay channel $K^{\ast }(892)K$ is also important $\left(
\Gamma =103MeV\right) ;$ the other channels are about $10MeV$. The total
decay width is around $600MeV$.

Table 4 give the results:

\begin{center}
\begin{tabular}{|c|c|c|c|c|c|}
\hline
$\Gamma _{\rho \left( 770\right) \pi }$ & $\Gamma _{\rho \left( 770\right)
\omega }$ & $\Gamma _{\rho \left( 1450\right) \pi }$ & $\Gamma _{K^{\ast
}(892)K^{-}}$ & $\Gamma _{K^{\ast }(1410)K^{-}}$ & $\Gamma _{tot}\left(
2.0\right) $ \\ \hline
$296$ & $100$ & $75$ & $103$ & $7$ & $581$ \\ \hline
\end{tabular}

\textit{Table 4: Predicted widths in }$\alpha _{s}MeV$\textit{\ for the
decay of a QE }$1^{-+}$\textit{-hybrid meson of mass }$2.0GeV.$
\end{center}

\subsection{The Gluon-Excited mode, decaying in $b_{1}\protect\pi $}

Concerning the GE mode, we find three hybrid states with an orbital
excitation between the gluon and the $q\overline{q}$ system, with:

\[
l_{g}=1,\ l_{q\overline{q}}=0,\ S_{q\overline{q}}=1,\text{ } 
\]

and respectively for the three states:

\[
J_{g}=L=0,1,2. 
\]

For an hybrid of mass $2.0GeV$, only 
\begin{eqnarray*}
&&f_{1}\left( 1520\right) \pi ^{-},\ f_{1}\left( 1420\right) \pi ^{-},\
f_{1}\left( 1285\right) \pi ^{-},\ b_{1\left( 1235\right) }\pi ^{-},\
a_{1}\left( 1260\right) \eta ,\  \\
&&K_{1}^{0}\left( 1400\right) K^{-},\ K_{1}^{-}\left( 1400\right) K^{0},\
K_{1}^{0}\left( 1270\right) K^{-}\text{ and }K_{1}^{-}\left( 1270\right)
K^{0}\text{ }
\end{eqnarray*}

final states are allowed on general grounds: conservation of angular
momentum, parity, isospin and G-parity. All this channels are allowed by the
selection rule for $1^{-+}$ GE-hybrid meson which decays only into one
fundamental meson and one orbitally excited meson $\left( L=0+L=1\right) $%
\cite{Iddir}$.$

The quark-gluon constituent model predictions for the decay of a $2.0GeV\ \
1^{-+}$ GE-hybrid meson are given in Table 5:

\begin{center}
\begin{tabular}{|c|c|c|c|}
\hline
$L$ & $0$ & $1$ & $2$ \\ \hline
$\Gamma _{b_{1\left( 1235\right) }\pi }$ & $166$ & $499$ & $832$ \\ \hline
$\Gamma _{a_{1}\left( 1260\right) \eta }$ & $156$ & $118$ & $196$ \\ \hline
$\Gamma _{f_{1}\left( 1520\right) \pi }$ & $118$ & $88$ & $148$ \\ \hline
$\Gamma _{f_{1}\left( 1420\right) \pi }$ & $142$ & $106$ & $177$ \\ \hline
$\Gamma _{\ f_{1}\left( 1285\right) \pi }$ & $160$ & $120$ & $202$ \\ \hline
$\Gamma _{K_{1}^{0}\left( 1400\right) K^{-}}$ & $119$ & $89$ & $149$ \\ 
\hline
$\Gamma _{K_{1}^{-}\left( 1400\right) K^{0}}$ & $117$ & $88$ & $147$ \\ 
\hline
$\Gamma _{K_{1}^{-}\left( 1270\right) K^{0}\text{ }}$ & $175$ & $131$ & $219$
\\ \hline
$\Gamma _{K_{1}^{0}\left( 1270\right) K^{-}}$ & $119$ & $89$ & $149$ \\ 
\hline
$\Gamma _{tot}\left( 2.0\right) $ & $1272$ & $1328$ & $2219$ \\ \hline
\end{tabular}

\textit{Table 5: Predicted widths in }$\alpha _{s}MeV$\textit{\ of a GE }$%
1^{-+}$\textit{-hybrid meson of mass }$2.0\ GeV.$
\end{center}

\section{\protect\bigskip The experimental data}

$i)$Hybrid $1^{-+}$ candidates (with masses $1400MeV$, $1600MeV$, $1740MeV$, 
$2.0GeV$) have been claimed to be observed\cite{Thomp, Black}, decaying in $%
\eta \pi $, $\eta ^{\prime }\pi $, $f_{1}\pi $ and $\rho \pi $:

\textbf{with masses around 1400\ MeV:}

\begin{itemize}
\item by BNL (in $\pi N$ interactions at $18GeV$) at $\left( 1370\pm
16\right) MeV$, with decay width $\Gamma =\left( 385\pm 40\right) MeV$,
decaying in $\eta \pi $, $\eta ^{\prime }\pi $ and $f_{1}\pi $\cite{Thomp,
Chung}.

\item by Cristall Barrel Coll. at LEAR (in $\overline{p}n$ interactions),
with a mass $\left( 1400\pm 20\right) MeV$, and $\Gamma =\left( 310\pm
50\right) MeV$, decaying in $\eta \pi $\cite{Abele}.

\item by GAMS Coll. at CERN, (in $\rho \pi $ interactions at $100GeV$), with
a mass $1400MeV$, and $\Gamma =180MeV$, decaying in $\eta \pi $\cite{Alde}$.$

\item by VES Coll. with mass $1405MeV$,\ decaying in $\eta \pi $ and $\eta
^{\prime }\pi $\cite{Bela, Proko}.
\end{itemize}

\textbf{with mass 1600\ MeV:}

\begin{itemize}
\item by E852 at BNL with mass $\left( 1593\pm 8\right) MeV$ and $\Gamma
=\left( 168\pm 20\right) MeV$, decaying in $\rho \pi $\cite{Adams, Chung}.

\item by E852 at BNL with mass $\left( 1597\pm 10\right) MeV$ and $\Gamma
=\left( 340\pm 40\right) MeV$, decaying in $\eta ^{\prime }\pi $\cite{Ivanov}%
.
\end{itemize}

\textbf{with mass 1740 MeV:}

\begin{itemize}
\item at FNAL, with mass $\left( 1740\pm 12\right) MeV$ and $\Gamma =\left(
136\pm 30\right) MeV$, decaying in $f_{1}\pi $\cite{Black}.
\end{itemize}

\textbf{with mass 2.0 GeV:}

\begin{itemize}
\item by AGS Coll. at BNL, decaying in $f_{1}\pi $ and $\eta \left(
1295\right) \pi $\cite{Lee}.
\end{itemize}

$ii)$The analysis of $e^{+}e^{-}$ data \cite{Donna} shows the possible
existence of $J^{PC}=1^{--}$states beyond the na\"{\i}ve \ quark model, like
a system of mixing of a $q\overline{q}g$ hybrid meson with $q\overline{q}$
state (for the $1.1GeV$ state).

$iii)$A suggestion to explain the $\Psi $ anomaly observed at CDF is by the
mixing of an hybrid meson with a mass around $4.1GeV$ and a conventional $%
\Psi \left( 3S\right) $ charmonium; the hadronic decays of the states $\Psi
\left( 4040\right) $ and $\Psi \left( 4160\right) $ are predicted dominated
by the $\Psi \left( 3S\right) $\ component\cite{Page}. We have studied the
states of the $1^{--}c\overline{c}g$ charmonium hybrid with mass around $%
4GeV $ , and the possibility to mixing with the conventional $c\overline{c}$
charmonium meson\cite{Safir1}.

\section{Theoretical results and comparison}

We recapitulate the results of the predicted $1^{-+}$hybrid mesons in Table
6; we pay our attention on the 1.32-2.04 GeV mass range (corresponding to
the mass range of the experimental candidates). We give the predicted masses
with the corresponding modes (QE or GE) and the favourite decay channels.

\begin{center}
\begin{tabular}{|c|c|c|}
\hline
mass (in GeV) & mode & favourite decay channels \\ \hline
$%
\begin{array}{c}
1.32 \\ 
1.34 \\ 
1.45%
\end{array}%
$ & $%
\begin{array}{c}
\text{QE} \\ 
\text{mixing dominated by QE} \\ 
\text{QE}%
\end{array}%
$ & $\rho \pi ,K^{\ast }K$ \\ \hline
$%
\begin{array}{c}
1.56 \\ 
1.58 \\ 
1.6%
\end{array}%
$ & $%
\begin{array}{c}
\text{GE} \\ 
\text{QE} \\ 
\text{mixing dominated by QE}%
\end{array}%
$ & $%
\begin{array}{c}
b_{1}\pi ,f_{1}\pi \\ 
\rho \pi ,K^{\ast }K,\rho \omega \\ 
\rho \pi ,K^{\ast }K,\rho \omega%
\end{array}%
$ \\ \hline
$%
\begin{array}{c}
1.69 \\ 
1.72 \\ 
1.72 \\ 
1.75%
\end{array}%
$ & $%
\begin{array}{c}
\text{GE} \\ 
\text{GE} \\ 
\text{mixing dominated by GE} \\ 
\text{GE}%
\end{array}%
$ & $b_{1}\pi ,f_{1}\pi $ \\ \hline
$%
\begin{array}{c}
1.84 \\ 
1.87 \\ 
1.89 \\ 
1.99 \\ 
2.02 \\ 
2.04%
\end{array}%
$ & $%
\begin{array}{c}
\text{GE} \\ 
\text{GE} \\ 
\text{GE} \\ 
\text{GE} \\ 
\text{mixing dominated by GE} \\ 
\text{GE}%
\end{array}%
$ & $b_{1}\pi ,a_{1}\eta ,f_{1}\pi ,K_{1}K$ \\ \hline
\end{tabular}

\textit{Table 6: Predicted hybrid }$1^{-+}$\textit{\ mesons with preferred
decay channels.}
\end{center}

Then, around the mass 1.4 GeV, the hybrid $1^{-+}$ mesons are predicted in
the QE-mode, or in a mixed (QE+GE)-mode, dominated by QE ($\theta _{\max
}\sim 4.6%
%TCIMACRO{\U{b0}}%
%BeginExpansion
{{}^\circ}%
%EndExpansion
$, $\lambda ^{2}\sim 0.006$); and the decay should occur into $\rho \pi
,K^{\ast }K$.

Around 1.6 GeV, we may obtain hybrid mesons in the two modes, decaying
preferably into $b_{1}\pi $ for the GE-hybrid, and $\rho \pi $ for the
QE-hybrid.

Around 1.7 GeV, the predicted hybrid $1^{-+}$ mesons will be built in pur
GE-mode, or in a dominated by GE mode; the decay should occur preferably in $%
b_{1}\pi $.

Around 2.0 GeV, we obtain pure GE-mode hybrid $1^{-+}$ mesons, or dominated
by GE; the preferred decay channels are into $b_{1}\pi $, or $K_{1}K$.

In our model, considering the different possibilities to construct a $%
J^{PC}=1^{-+}$ hybrid meson, relative to the two modes (GE and QE), we
obtain the following predictions for candidates with masses $1.4,\ 1.6,\ 1.7$
and $2.0GeV$ , compared with the experimental data (Table 7):

\begin{center}
\begin{tabular}{|c|c|c|c|c|}
\hline
mass (in $GeV$) & mode & channels & decay widths (in $MeV$) & exp.data \\ 
\hline
1.4 & QE & $\rho \pi $ & $72$ & - \\ \hline
1.6 & QE & $\rho \pi $ & $142$ & $168$ \\ \hline
1.6 & GE & $%
\begin{array}{c}
b_{1}\pi \\ 
f_{1}\left( 1285\right) \pi \\ 
f_{1}\left( 1420\right) \pi%
\end{array}
$ & $%
\begin{tabular}{c}
$230$ \\ 
$186$ \\ 
$62$%
\end{tabular}
$ & $%
\begin{array}{c}
\text{-} \\ 
\text{seen} \\ 
\text{-}%
\end{array}
$ \\ \hline
1.7 & GE & $%
\begin{array}{c}
b_{1}\pi \\ 
f_{1}\left( 1285\right) \pi \\ 
f_{1}\left( 1420\right) \pi%
\end{array}
$ & $%
\begin{tabular}{c}
$231$ \\ 
$188$ \\ 
$62$%
\end{tabular}
$ & $%
\begin{array}{c}
\text{-} \\ 
136 \\ 
\text{-}%
\end{array}
$ \\ \hline
2.0 & GE & $%
\begin{array}{c}
K_{1}K \\ 
b_{1}\pi \\ 
f_{1}\pi \\ 
a_{1}\eta%
\end{array}
$ & $%
\begin{array}{c}
175 \\ 
166 \\ 
160 \\ 
156%
\end{array}
$ & $%
\begin{array}{c}
\text{-} \\ 
\text{-} \\ 
observed \\ 
\text{-}%
\end{array}
$ \\ \hline
\end{tabular}

\textit{Table 7: Predicted decay channels and rates for hybrid candidates
with masses }$1.4,\ 1.6,\ 1.7$\textit{\ and }$2.0GeV.$
\end{center}

Note that for the GE-mode, we have reported the results only for L=$0.$

PS: the results for the $J^{PC}=1^{-+}$ hybrid mesons at 1.4 and 1.6 GeV are
given from \cite{Safir2}.

Table 8 presents our total decay widths predictions for $1.4,\ 1.6,$\textit{%
\ }and\textit{\ }$2.0GeV\ $hybrid candidates.

\begin{center}
\begin{tabular}{|c|c|c|c|}
\hline
Total decay width & L=0 & L=1 & L=2 \\ \hline
$\Gamma _{QE}\left( 1.4\right) $ & - & 72.3 & - \\ \hline
$\Gamma _{QE}\left( 1.6\right) $ & - & 165 & - \\ \hline
$\Gamma _{GE}\left( 1.6\right) $ & 708 & 1564 & 2606 \\ \hline
$\Gamma _{GE}\left( 2.0\right) $ & 1272 & 1328 & 2219 \\ \hline
\end{tabular}

\textit{Table 8: The total decay widths of the }$1^{-+}$\textit{\ hybrid
mesons at }$1.4,\ 1.6,$\textit{\ and }$2.0GeV\ .$
\end{center}

In our model, the gluon-excited $1^{-+}$ hybrid meson do decay preferably in 
$b_{1}\pi $; this result was confirmed by the flux-tube model \cite{Donna-pa}%
. The quarks-excited $1^{-+}$ hybrid meson do decay preferably into $\rho
\pi $.

The decay channels into $\eta \pi $, $\eta ^{\prime }\pi $, $\eta (1295)\pi $
are forbidden, both in the two modes: in the GE-mode, due to the selection
rule which suppresses this channel (and allows a channel with one
fundamental meson and the other one orbitally excited: (L=0)+(L=1)); in the
QE-mode, due to the total spin conservation in the final state. This is in
contradiction with experiment.

The total decay widths of the 1.6 (in the GE-mode) and 2.0 GeV $1^{-+}$
hybrid meson are found to be very large, which means that such states do not
really emerge from the continuum of the two meson spectrum. In other words,
the GE-hybrids in the 1.6-2.0 GeV mass range do decay before they had time
to really exist as hadrons: \textit{they do not exist as resonances}.

On the contrary, the QE -1.6 candidate has a total decay width around 165
MeV, with a $\rho \pi $ preferred decay channel, in agreement with BNL. Our
conclusion is that \textit{this resonance may be considered as a hybrid
meson in the quarks-excited mode.}

\section{Discussion and Conclusion}

We have presented here some theoretical results and experimental data
relative to exotic mesons, which are possibly interpreted as hybrid $q%
\overline{q}g$\ candidates. We used the quark-gluon constituent model,
implemented with a confining potential between the hybrid constituents,
which is inspired from gluon-exchange in QCD. However, the flux-tube model
assumes that hybrids are predominately quark-antiquark states moving on an
adiabatic surface generated by an excited color flux-tube; there is no
constituent gluon in FTM model, the gluon degrees of freedom are treated as
collective excitations of the color flux; in practice, the color is treated
as a string, and the excited states are represented by the excited modes of
the string.

Restricting ourselves to the lightest states, we have three hybrids (L=0, 1,
2) with an orbital excitation between the gluon and the $q\overline{q}$\
system (GE-mode), and one with an orbital excitation between the $q$ and the 
$\overline{q}$ (QE-mode). Mixed (QE+GE) states may exist, but the results
show that the QE-hybrid and the GE-hybrid mix very weakly.The QE-mode is
lighter than the GE-one (see Table 1, Table 2), due to the confining
interaction inside each mode.

In our model, the GE-hybrid and the QE-hybrid are allowed to decay
respectively into two mesons S+P states and into two ground state mesons.

The $\left( b_{1}\pi \right) $ and $\left( f_{1}\pi \right) $\ are the
preferred decay channels of the hybrid mesons in the GE-mode; the first
channel was confirmed by the flux-tube model \cite{Donna-pa}, but \textit{%
was never observed}, and it will be necessary to check this channel in order
to verify our selection rule concerning the GE-mode. The latter channel was
observed by AGS\ Coll. at BNL\cite{Lee}, in the 1.6-2.2 GeV mass range, but
without precision on the corresponding branching ratio.

The $\left( \rho \pi \right) $ is the favourite decay channel for the
QE-mode and was observed only for the 1.6 GeV resonance by E852 Coll. at BNL 
\cite{Adams}, with a width $\Gamma =\left( 168\pm 20\right) MeV$, which is
in good agreement with our prediction $\sim $142 MeV.

The \ $\eta \pi $, $\eta ^{\prime }\pi $, $\eta (1295)\pi $\ channels are
forbidden for every resonances and both in the two-modes (QE and GE). 
\textit{They have been observed}. Unhappily, it becomes difficult to
understand the $\sim $400 MeV width\ of the E852 Coll.\cite{Thomp} as an
evidence for an exotic resonance at 1.4 GeV, which is totally forbidden in
our model.

Theoretical studies by QCD-Sum Rules give $\left( \rho \pi \right) $\ as
dominant decay channel compared to $\eta \pi $ and $\eta ^{\prime }\pi $\cite%
{S.R, Zhang}.

The total decay width of the $1.4GeV\ 1^{-+}$ hybrid, dominated by the
QE-mode, is around 72 GeV, and the decay channel is into $\rho \pi $
preferably.

The total decay widths of the GE-1.6 and 2.0 GeV hybrids are respectively
around 700 MeV and 1270 MeV for $L=0$ and even larger for $L\geq 1$. The
increase of the widths with the resonance mass is obviously due to the
increasing phase space. Then\textit{\ the GE-1.6 and the 2.0 GeV hybrids do
not exist as resonances.}

Ref\cite{Donna-pa} rejects the hypothesis that the observed 1.4 and 1.6 GeV
states are both hybrid mesons, because of the experimental width which is
larger for the 1.4 state than the 1.6 state, which should be opposite. They
interpret the experimental observed peak in the $\eta \pi $ channel, which
is suppressed by symmetrization selection rules, as a sizable final state
interaction and they suggest that the E852 $\eta \pi $\ peak is due to the
interference of a Deck-type background with an hybrid resonance of higher
mass, for which the $\widetilde{\rho }$ at 1.6 GeV is an obvious candidate.

Then from our analysis, we conclude the following:

\begin{enumerate}
\item The $1^{-+}$ resonance at $1.4GeV$ would be a good hybrid candidate if
it has been observed in $\rho \pi $, which is predicted around 70 MeV.
However, it is unplausible to interpret the $\sim $400 MeV width in $\eta
\pi $ \ reported by E852 experiment.

\item The $1^{-+}$ resonance at $1.6GeV$ is a good hybrid candidate in the
QE-mode, with a total decay width around 165 MeV dominated by $\rho \pi $,
in good agreement with BNL.

\item For the $1^{-+}$ resonance at $2.0GeV$, dominated by the GE-mode, it
has a total decay width around 1.3 GeV for L=0, and even larger for L$\geq 1$%
, making the hybrid interpretation rather unbelievable.
\end{enumerate}

Finally, to check the existence of the hybrid mesons, namely the $1^{-+}$\
ones, and have more informations about the physics of these objects, it will
be important to have (from the experiments) the observed branching ratio in $%
f_{1}\pi $\ in the $1.6-2.2GeV$ mass range, and to check the $b_{1}\pi $\
channel. Concerning the $1.4GeV$, it will be necessary to check the $\rho
\pi $\ channel. We believe that these experimental tests are crucial to
remove any doubt concerning the interpretation of these objects and to lead
to the understanding of the hybrid phenomenology.

\begin{quote}
\textbf{Aknowledgments:}
\end{quote}

We would like to thank A.S.Safir \emph{(Univ. of Muenchen)} for his
contribution to this work. We are grateful to O. P\`{e}ne \emph{(Laboratoire
de Physique Th\'{e}orique, Univ. of Paris-sud)} for extremely useful
discussions.

\end{document}